\documentclass[12pt]{article}
\usepackage{amsmath, amsfonts, amssymb}
\usepackage{amsthm}
\theoremstyle{remark}

\newcommand{\beq}{\begin{equation}}
\newcommand{\eeq}{\end{equation}}
\newcommand{\beqa}{\begin{eqnarray}}
\newcommand{\eeqa}{\end{eqnarray}}

\newcommand{\ket}[1]{| #1 \rangle}

\newcommand{\hbcommentoff}[1]{{}}
\begin{document}

\begin{center}{\bf Quantum knowledge, quantum belief, quantum reality}\\
{\bf Notes of a QBist fellow traveler}\\
Howard Barnum  \\October 7--16, 2009 \end{center}

\begin{abstract}
I consider the ``Quantum Bayesian'' view of quantum theory as
expounded in a 2006 paper of Caves, Fuchs, and Schack.  I argue that
one can accept a generally personalist, decision-theoretic view of
probability, including probability as manifested in quantum physics,
while nevertheless accepting that in some situations, including some
in quantum physics, probabilities may in a useful sense be thought of
as objectively correct.  This includes situations in which the
ascription of a quantum state should be thought of as objectively
correct.  I argue that this does not cause any \emph{prima facie}
objectionable sort of action at a distance, though it may involve
adopting the attitude that certain dispositional properties of things
are not ``localized'' at those things.  Whether this insouciant view
of nonlocality and objectivity can survive more detailed analysis is a
matter for further investigation.
\end{abstract}

\section{Introduction}
Quantum states are the heart of the predictive and explanatory power
of quantum physics.  On any interpretation of quantum theory, the
quantum state (wavefunction or, more generally, density operator) of a
system is a tool for calculating the probabilities of whatever future
measurements we may choose to make on the system.  In fact, the
quantum state can be considered to be a succinct \emph{representation}
of the probabilities for the outcomes of any possible measurement on
the system.  In this paper, I consider interpretations of quantum
theory on which this is the \emph{essence} of quantum states, and
focus on the nature, objectivity, and reality of the probability
ascriptions represented by quantum states.  Since I subscribe to a
broadly Bayesian, personalist account of probability insofar as it
relates to the world, my analysis will have to make sense of quantum
probabilities in personalist Bayesian terms.  But the main concern of
the paper will be to argue that a notion of objective
probability---``propensity'', if you like---can be compatible with
the personalist view, and that quantum probabilities sometimes have
this nature.

More specifically, while I agree with the ``Quantum Bayesian'' or
``QBist'' perspective of Carlton Caves, Christopher Fuchs, and
R\"udiger Schack as expressed in \cite{CFS} on many points, in this
paper I will take---as have others---a position which some (including CFS, 
an acronym which I will use to refer to Caves, Fuchs, and Schack as authors
of \cite{CFS}; some of their views may have changed since then)
might consider more ``straight Copenhagen''.  The position is that in
at least some situations an ascription of a quantum state to a system
can be ``objectively correct'' in light of quantum theory, general
background knowledge and common sense, and specifics concerning the
system in question (how it was prepared, the outcomes of previous
measurements made on it, and so forth).

Having argued for this claim, I will then examine what it implies about
the nature of reality, and of the quantum state's relation to it,
especially in light of the arguments of EPR and the evident concern of
the ``QBists'' about a link between objective probabilities and
objectionable nonlocal effects.  I will argue it does not support any
claim that the quantum state is a real \emph{entity}, and although it
may sometimes commit us to ascribing dispositional or modal
\emph{properties} to systems, which are perhaps real in the weak sense
that systems \emph{really have them}, it is far from clear that we
must view these modal properties as \emph{located in} the system, or
that their nonlocal change when the state of a system ``collapses''
when a measurement result is obtained on its entangled partner system,
should bother us.

\section{Personalist probability and quantum states}

\subsection{This I believe: personalist Bayesianism is the essence of applied probability}
Let me re-emphasize that I view probabilities that occur in physical
theories as ultimately guides to action.  They determine rational
betting odds, if you like.  Personalist, decision-theoretic accounts
of probability, along the lines of Bruno de Finetti, Leonard Savage
\cite{Savage}, and more recently, Richard Jeffreys \cite{Jeffrey}, are just fine
with me.  The point of including Jeffreys is to emphasize that I'm not
wedded to {\em conditioning} on newly acquired certain knowledge as
the only reasonable way in probabilities can change in light of our
interactions with the world, a point on which I agree with CFS.
Rather, maintaining {\em coherence} of one's probabilistic beliefs in
the light of continued interactions is the goal.  Let me also
emphasize---for it will be important---that I view \emph{scientific
  progress} itself as best understood---partly in the sense of
``rational reconstruction'' but also to a large extent as regards
actual, if often implicit, practice (cf. \cite{Jeffrey})---in terms
of such Bayesian (broadly-speaking) updating of beliefs.  As an
undergraduate studying philosophy at a time when, and in a department
where, Nelson Goodman's \emph{Fact, Fiction, and Forecast}
\cite{Goodman} and Quine's ``Two dogmas of empiricism'' \cite{Quine}
were iconic reference points, I took the lesson of Goodman's paradox
of the grue emeralds to be the inescapable need for prior beliefs
about the plausibility of alternative hypotheses compatible with the
data, in order to get any kind of inductive reasoning off the ground,
and soon adopted a generally Bayesian analysis of ``inductive
reasoning''.  The Duhem-Quine thesis that our beliefs meet the world
as a body, a linked web which may adjust to empirical experience in
various ways also formed part of this view.  This makes me suspicious
of views which rely too heavily on ``foundational'' approaches to
knowledge and scientific progress in our corpus of beliefs contains a
sharply delineated foundation of ``certain knowledge'', for example
the ``protokolsatze'' of some of the early Vienna School positivists.
Still, within the ``web of belief'' some statements may be closer to 
our immediate experience than others, so I am wary of adopting a too-glib
anti-foundationalism as well. 
A sympathetic view of the Quine-Duhem thesis also disposes me toward
the Richard Jeffreysian view of of belief change while maintaining
probabilistic coherence which Fuchs and Schack have recently
recommended to me.  This broadly ``Bayesian'' viewpoint is still at the core
of my views on knowledge and how we acquire it.

\subsection{Quantum probabilities, personalist and objective}

It might seem surprising, then, that I would favor the view that
quantum probabilities can be, in important situations, objective.
This should not, however, be so surprising: by ``objective'' I do not
mean that they are not to be understood in personalist, Bayesian terms
as ``guides to action'', nor do I mean they are facts about the
frequencies of certain types of events in a ``block world'' in which
all events, past and future, are laid out, described, and classified.
I mean that in the sense in which the fact that the Sun will rise
tomorrow is an \emph{objective fact} of physical and astronomical
science, I'd grant objectivity to some quantum probabilities.  That
the Sun will rise tomorrow is a proposition one could logically
disbelieve; no evidence we currently have \emph{logically compels} one
to believe it, and one could presumably come up with priors under
which the currently available evidence would still leave one in
significant doubt about it.  It is, in the end, a personalist
probability judgement of near-certainty for a particular event---a
judgement that virtually all of us agree on, and a judgement that we
view as objectively correct.  It is in this sense that I want to
maintain that quantum states can be objectively correct, as a matter
of quantum physics plus background knowledge.  While at some deeper
level it---like all judgements---may be in some sense personalist and
subjective, I believe that for one who accepts quantum
physics, plus various uncontroversial statements about observable
(``classical'' if you like) facts concerning experimental setups,
there are situations in which there's a particular state vector or
density matrix that represents the correct beliefs, and that should guide
one's future actions, concerning this quantum system.

The view I defend here may be generally identified with what CFS call
``Copenhagen-like interpretations'' of quantum mechanics, but there is
one possibly critical difference.  CFS say ``To our knowledge, these
[``realist readings of the Copenhagen interpretation'']\footnote{By 
``realist readings'' I take CFS to mean readings that are realist about 
``the world'', not necessarily realist about the quantum state} all have in
common that a system's quantum state is determined by a sufficiently
detailed, agent-independent \emph{classical description} of the
preparation device, which is iself thought of as an agent-independent
physical system.''  I would not want to claim that this holds in all
situations.  The terminology ``the preparation device'' suggests that
CFS only intend it to hold when a system is viewed as ``having been
prepared'', but I'm not sure I'd even go that far.  Nevertheless, I'd
claim, perhaps contra CFS, that in many such preparation situations,
the quantum state is determined (not in the causal sense) by
``classical'' facts about preparation devices.  I don't necessarily
take ``classical'' in the sense of ``classical physics'';
``ordinary-language'' might be closer.  But the description could
contain technical, theory-laden terms, like ``polarizing
beamsplitter''; see below.

It's not always clear to me that this represents a serious divergence
from CFS' views, rather than a matter of emphasis.  

It's important here to separate, at least \emph{a priori}, the issues
of the \emph{objectivity} of quantum states, from issues of their
\emph{reality} or ontological status.  The term ``objectivity'' with its
root of ``object'' suggests an assignment of ontological status to ``objective
probabilities'' which I'd like the option of avoiding unless it becomes 
clear that it's compulsory.

Another point that is important to CFS is the agent-dependence of
quantum states.  Fuchs and Schack like to emphasize that the quantum
state is relative to a specification of an agent and system.  When I
do an experiment on $X$, $X$ has a wavefunction for me (or I have one
for it).  On a more objective-probabilities view quantum theory
sometimes tells me what my wavefunction should be.  But this could, if
needed, be understood as advice for a certain agent, contemplating a
certain thing ``as quantum''.  I am willing to accept a degree of
agent-dependence---or, perhaps, background-information dependence---of
quantum probabilities even in cases when I believe there may be
objective correctness to these probability assignments.

In my view, ``objective correctness'' is not exactly a matter of
rejecting the notion that these probabilities ``always depend on a
prior.''  Rather I believe that just about everything in science and
everyday life ``depends on a prior'' if ``rationally reconstructed'';
but that in many things, there is sufficient agreement that we have
converged on subjective probabilities that they may as well be granted
the term ``objective''.  In particular, even the everyday-life
assertions of fact that are taken as unproblematic, I take to have
something like this character.

Perhaps my notion of objective probabilities is so subjective that,
CFS would be willing to endorse it.  In this case, one wonders why
they go to such great lengths to emphasize the subjective nature of
quantum probabilities.  One likely reason is that they wish to make it
clear that they do not accept the notion of an ``objectively real''
quantum state as physical object, of the sort that appears in most
versions of the de Broglie-Bohm and Everett interpretations.  But I
worry that they are underestimating the extent to which we can have
both all the objectivity anyone could reasonably want for at least
some quantum state ascriptions---a degree of objectivity which,
moreover, it seems to me would be unreasonable not to grant in some
cases---while also not {\em reifying} the quantum state as an object
out in the world, and especially, not as an object localized at the
system whose state it is.  I believe that an unfounded fear that these
two features of an interpretation might be incompatible, and of the
``action at a distance'' that would follow from dropping the latter
feature and allowing quantum states as physical objects in the world,
has led CFS to project an excessively subjectivist view of the nature
of quantum theory, and to unnecessarily differentiate their views from
that part of the Copenhagen tradition---represented in our time by,
for example, Mermin, Brukner and Zeilinger, and earlier by Peierls and
perhaps even Bohr---that views the quantum state as essentially
epistemic, in the sense of representing our best guide to action in
the face of uncertainty about the outcomes of alternative possible
measurements.

There is more to it, however: I believe that CFS at different places
adopt two versions of the claim that quantum theory never determines
state ascriptions, corresponding to two different interpretations of
the words ``quantum theory'' in this claim.  The stronger version,
which I take exception to, claims that ``quantum physics'' broadly
construed never determines state ascriptions; the weaker one, which
seems obviously correct, claims that ``the quantum formalism'' of
density operators and positive operator valued measures does not
determine it.  There are discussed at more length in the next section, 
and the implications of the distinction for the QBist program, and related
programs viewing the quantum state as epistemic, are discussed to in the 
conclusion.

\section{Meaty quantum physics versus \emph{th\'eorie quantique minceur}}

Before going into more detail about objective state assignments and
state preparation, I make a distinction which bears importantly on the
meaning of ``Quantum Bayesianism'' and on the nature of the QBist
project (as I imperfectly understand it).

The distinction is between \emph{quantum physics} and what might be
called \emph{the quantum formalism}, or more specifically, the Hilbert
space/POVM formalism.  The quantum formalism is what is sometimes
called ``quantum probability'' (not to be confused with notions of
not-necessarily-positive real or complex-valued probabilities that
sometimes go by this name): a particular generalization or
``extension'' of the formal apparatus of classical Kolmogorovian
probability theory, from among the many mathematically consistent ways
of extending this theory to allow for the possibility of choosing
alternative, incompatible measurements (or equivalently, incompatible
random variables).  Such extensions model a system by specifying a
compact convex set of possible normalized system states, which can
serve as the base for a cone of unnormalized states, and a compact
convex set of possible measurement outcomes that is an initial
interval in the cone dual to the cone of states (or at least, in a
subcone of this)\footnote{See, for example, \cite{Edwards} or the
  introductory sections of \cite{BBLW06,BBLW08}, among many
  possible references, for introductions to this formalism.}.  There
is nothing nonstandard about the probabilities that occur in such
models; mathematically, they are completely standard and therefore the
states in such models are perfectly susceptible to Bayesian
interpretation as guides to betting on the outcomes of measurements.
The quantum formalism, the formalism of Hilbert space density matrices
(the normalized states) and POVM elements (the initial interval in the
dual cone), is a specific case of this general formalism.  The
principle that the states one can prepare, and the measurement
outcomes one can obtain, for a system, can be modeled within this
quantum formalism is indeed, therefore, ``an empirical addition to
Bayesian coherence''.  I'd add only that the general framework of
convex state and effect spaces---as further developed to include
dynamics (both unconditional and conditional on measurement results),
can be understood as representing ``Bayesian coherence'' in settings
where alternative, ``incompatible'' (in a sense that can be made
precise within the convex sets framework) measurements may be made on
a given system.

\emph{Quantum physics}, on the other hand, is the full, meaty physical
theory, in which the states in particular orthonormal bases carry
physical significant labels, and we truck with notions of forces,
particles, interaction Hamiltonians, fields, and so forth.

The line between meaty quantum physics and the math-y quantum
formalism (the ``th\'eorie quantique minceur'' of the section title)
is somewhat blurry.  Preferred tensor factorizations,
observables, symmetries might help make the transition between the
two.  Indeed, symmetries may lead us to consider convex state spaces
other than the most standard quantum ones, for instance through Abelian or
nonabelian superselection rules.  

An important point is that the quantum formalism may be viewed as a
\emph{representation} of probabilities, for example for particular
preparation procedures followed by particular results of particular
measurement procedures---arrived at through the usual process of
scientific inference, which I view as in essence Bayesian.  Such
actual preparation and measurement procedures contain plenty of meaty
detail in their description. A formal probabilistic structure can be
abstracted from them---for example, by the time-honored strategy of
lumping together as ``preparing the same state'' all preparation
procedures that lead to the same list of probabilities of the various
measurement outcomes' occuring, and lumping together as ``the same
effect'' all measurement outcomes that give rise to the same list of
probabilities of occuring in the various states.  But it should not be
forgotten that this structure has its roots in such experimental
reality, and thus, indeed, in the very same inferences that led us to
adopt the full ``meaty'', ``physics-y'' quantum theory itself.

Statments like that of Fuchs and Schack, that ``the basis for one's
particular quantum state-assignment is always \emph{outside} the
formal apparatus of quantum mechanics'' suggest that they may only be
concerned to assert subjectivity relative to ``the quantum
formalism'', and not to ``meaty quantum theory'' as I've described it
above.  This would provide a way out of the need to deny the ``objectivity'' 
of quantum state ascriptions such as that of the state prepared by a
polarizing beamsplitter, etc...  These state-ascriptions are objective
according to \emph{quantum physics}, but are certainly not determined by 
the mere \emph{quantum formalism}.  I am not sure if this is really the 
route CFS, or at least Fuchs and Schack, want to take;  if so, I think 
we are close to complete agreement, but in \cite{CFS} there still seems a push
to deny the objectivity even of state assignments warranted by full, meaty
applied quantum physics.  

\section{State preparation}
Let's take an example where quantum probabilities may reasonably be
thought to be objective.  Send a single photon through a polarizing
beamsplitter and do high-efficiency photodetection at one output.
When the photon passes without a count, \emph{quantum physics} plus
general background knowledge determines the probabilities of counts at
the outputs of a second photodetector, perhaps differently oriented,
placed after the free output of the first---i.e. the state of the
polarization degree of freedom.

\subsection{Quantum description of the measurement device}
CFS argue against the contention that such a probability assignment may
be objective on the grounds that even in cases of state preparation, 
the fact that the preparation device carries out the claimed state preparation
operation, depends on subjective beliefs summarized in the quantum state of
the preparation apparatus.  

I agree that \emph{if one does} give a quantum description of the
measuring device, the operation it performs will be dependent upon
one's assignment of a quantum state to it---one's beliefs about it.
But who says a quantum analysis of the apparatus is needed?  From a
point of view from which what we take to be quantum depends upon our
purposes and the experiments we plan to do or other projects we plan
to undertake, to suddenly \emph{require} that we do a quantum analysis
of the apparatus seems to put us in danger of becoming unwitting
victims of the cult of the larger Hilbert space.  We don't need a
weatherman to say which way the wind blows, and we don't need a full
quantum analysis of something to tell us it's a polarizing
beamsplitter, though it is nice that a quantum analysis jives with a
more everyday story.  This observation is somewhat analogous to the
observation that we don't need neuroscience to (usually) trust in our
perceptions, except that ``polarizing beamsplitter'' is certainly more
theory-laden than many terms used in reporting perceptions.  I'd argue
that this theory-ladenness does not invalidate the claim that the
probabilistic description of the operations performed by such devices
can be viewed as developed to some extent at the phenomenological
level.  Indeed, these descriptions were developed as part of the
process that created quantum theory \emph{as a theory that gets its
  hooks into the physical world}.  In this process we already made
inferences to---ultimately personalistic, but about as objective as
anything gets in physical theories---probabilistic statements about
interactions between systems like that and measuring devices like
this.  These inferences were not themselves based on full-fledged
quantum theory, nor on explicit prior quantum state assignments to
measurement apparatus.  Rather, there was a simultaneous development
of concepts like ``photon'', ``photon polarization'', and ``polarizing
beamsplitter.''  If, upon making the optional but, it might be argued,
obligatory-that-it-be-in-principle-possible-to-make, transition to a
\emph{quantum} view of the measuring apparatus (or part of it), this
commits us to quantum state ascriptions, perhaps that is all right.
That is, after all, part of what I've been arguing is part of a
\emph{meaty} (or if you prefer, tofu-y or
texturized-vegetable-proteiny) version of quantum physics, which is
the kind we've actually got.  Of course in the end this probably won't
commit us to a pure state ascription, but only to specifying some very
coarse-grained features of the quantum state of the apparatus that are
sufficient for it to perform the operation it does.  This might be
reflected in assigning a highly mixed density operator with certain
properties---or even just asserting certain facts about the apparatus
density operator, without necessarily reflecting further on the exact
density operator we choose within the subspace of satisfactory ones.

\subsection{Estimating the behavior of apparatus}
Another important point is that we can do statistical estimation of
the relevant facts about how our apparatus works (and experimenters
often do), starting from an exchangeable prior for repeated use of the
beamsplitter that's not necessarily based on quantum theory.\footnote{
  An experimenter's implicit prior about the performance of an
  apparatus is likely to be much more complicated than just straight
  exchangeable ones---in particular, for a single complex experimental
  setup used many times in succession, they are likely to include a
  reasonable weight on the possibility of systematic, time-correlated
  drift due to environmental conditions in the lab; tighter
  correlations in a single ``run'' of use of the setup stemming from
  the effects of initializing the apparatus, and so forth.  Such
  considerations don't necessarily vitiate my basic point.}  Yes, as
CFS point out, this involves a prior; but many of our priors, such as
the likely exchangeability of certain sequences of events, reach
beyond the specifics of quantum theory into the way we think everyday
objects and situations behave.  Experimental apparatus can be
defective from time to time, or badly designed, and in some
cutting-edge situations there may be real questions about whether
quantum effects critical to its correct operation have been analyzed
correctly---but it would take some pretty kooky beliefs about quantum
states (or less outlandishly, some far-fetched paranoia about the
folks at the optics company) to lead to nonstandard beliefs about a
polarizing beamsplitter that would survive a few basic checks.

\section{Elements of reality?}

I've argued that sometimes it is an objective fact, in light of quantum physics and other
knowledge, what the probabilities of certain measurement outcomes
would be should we make the relevant measurement.  Does that make
these probabilities \emph{elements of reality?}

EPR famously said ``yes'', at least in the case when a measurement outcome will
give probability $1$.  It's worth including the passage:

``A comprehensive definition of reality is, however, unnecessary for
our purposes.  We shall be satisfied with the following criterion,
which we regard as reasonable.  \emph{If, without in any way
  disturbing a system, we can predict with certainty (i.e. with
  probability equal to unity) the value of a physical quantity, then
  there exists an element of physical reality corresponding to this
  physical quantity.} [...] Regarded not as a necessary, but merely as
a sufficient, criterion of reality, this criterion is in agreement
with classical as well as quantum ideas of reality.''

However, I suspect that probabilities can be objective without being elements of
reality.

Here's how I'd like to view the situation.  Quantum theory
\emph{recommends---or insists on} certain constraints on
decisionmaking, in some particular situations in which an agent knows
a certain measurement will be performed (or measurement-like situation
will arise), and has particular background knowledge.  Perhaps one
might want to say the quantum state is ``law-like''.  What I'd really
like to say, though, is that this fact about how one should bet
describes a {\em dispositional property} of the system---it is
disposed to give a certain outcome, if a particular measurement is
performed.  But is this property \emph{real}?  Well, what does that
even mean?  If we mean ``does the system really have the property?'',
well, sure.  But does the property \emph{exist}?  What are some
analogies that would help us understand what this means.  Is it like
the roughness of a rough rock?  I guess we could say the roughness of
the rock exists (I'm not sure how compelling that is) and seems to
be located at the rock.  But the roughness is a \emph{property}, not
an object.  Does that matter?  Well, even if we're led to say it's
``real'', that won't make it an \emph{object} as it is in the Everett
interpretation.  How would we treat objective probabilities in a
classical theory whose basic equations were stochastic equations
describing, say, the movements of particles?  Again, I think we'd say
they are lawlike statements about the dispositions of particles; we
wouldn't necessarily say the transition probabilities were themselves
an existent field.

Of course, some versions of Bohmianism say the quantum state
(``guiding wave'') is lawlike, rather than object-like.  My version of
the Copenhagen interpretation, though, isn't Bohmian for other
reasons: the probabilities are taken to concern, not underlying ontic
``positions'' and ``momenta'' but measurement outcomes.  Of course,
these outcomes may be taken to include outcomes \emph{of position
  measurements}.  But there are other observables that can be measured
as well, and the theory is silent on what's measured where.

Well, say we take it as a real property.  Should we be bothered by 
the prospect of its nonlocal change?

Are such properties even {\bf at the system}?  

Should EPR bother us?  Instantaneous change of a state of knowledge
needn't bother us.  Instantaneous communication would bother us.  But
this doesn't permit it.  Let's compare this to the real property of,
say, roughness.  We would be more worried if we could change something
from rough to smooth instantaneously at a distance, because that could
be used to signal.  The fact that the change in properties of $B$
represented by collapse in the EPR scenario \emph{cannot} be used to
signal seems to be a deep difference between such properties and more
commonplace ones like roughness.  Is it a manifestation of the
intrinsically relational, intrinsically less localized, nature of such
quantum properties? I guess I'd say so. (Or should we instead take
this as evidence that they are ``less real''.  Again I wonder what
would be the point of using, or not using, this form of words.  It is
probably better to attend to how the property actually behaves,
compared to e.g. roughness.)

We seem on solid ground if we wish to maintain that the fact about how
Alice should bet is not a fact about \emph{how things are at Bob's
  site}.  It's not reasonable to say that quantum physics recommends
that Bob should immediately change his betting behavior to the one it
recommends for Alice.  (Though I suppose it does claim that he'd be
better off if he did.)

The determinants of states can be ``unproblematic'', macro-facts.
E.g. in EPR, Alice's measurement results, even if relativistic
causality prevents them being known to Bob for awhile.

\subsection{Instrumentalism, reality, and the quantum state} The quantum state is
certainly a useful instrument even if it isn't a real property of the
system.  Emphasizing that a \emph{correct}, \emph{objective}, or
\emph{fact-like} quantum state is a correct (objective, fact-like)
prescription for betting---a good guide to action---stresses this useful,
instrumental nature of the state.  Does this give reason to doubt that
the such objective quantum probabilities are a ``real properties'' of
the world?  

It can be an \emph{objective fact} which instrument is useful in a
given situation.  But also, the fact that a certain instrument is
right to use in a given situation tells us something about how things
really are in that situation.  So an instrumental reading of the
quantum state doesn't necessarily get us out of viewing quantum
probabilities as ``real properties'' (or in more EPR-like language,
``corresponding to'' real properties).  Indeed, maybe more facts than
we think about reality are of this nature---enough so that we'll want
to call $\ket{\psi}$ real after all?.

``That's a table in front of me'' vs. ``reality is such that I'm
well-advised to use the concept of table, with all the predictive (it
will support things) and retrodictive (somebody made it) baggage it
brings along, in dealing with this particular situation.''  Not much
difference, is there?  Words as tools, as the cartoon version of
Wittgenstein goes.  Facts are stated using words.  To state a fact is
to use a tool?

Could the counterfactual (dispositional) nature of the property ``will
yield spin up in a measurement of $\sigma_x$'' save it from being
real?  But one might argue all, or many, concepts involve such
dispositional aspects.  

Indeed, the notion of reality might seem to {\em require} some
counterfactual thinking.  Does it necessarily describe agents who manipulate,
interact?  Hacking on realism about entities: ``if you can spray them, they
are real''.  We are, of course, more concerned with properties.  Is \emph{sprayability} 
real?

On balance, I'm inclined to think of quantum states, in situations
when quantum physics broadly construed dictates the state, as real
properties of the world.  I'm really not sure what ``real'' adds to
``property'' in this statement, though.  Should we think of them as
real properties of the systems whose states they are?  Again, what
does ``real'' add?  Does it mean they are \emph{not relational}, i.e.
they are independent of properties of other systems?  In this
case---depending on what is meant by ``independence''---the property
may not be real.  If ``independent'' means that it only concerns what
will happen in future measurements on the system, then perhaps the
property is not relational.  Of course, realizing the counterfactual
involves bringing it into relation with another system, the measuring
system. But as a disposition to behave a certain way if such relations
are brought about, it is not itself relational.  On the other hand, we
may, as in the EPR case, \emph{ascribe} the property because of
relations to other systems---so in \emph{that} sense it may need to be
thought of as relational.

There is no decisive test one can perform \emph{on a system, B}, to
determine whether $B$ is, or is not, in state $\ket{\psi}$.  There are
tests---measurements in a basis containing ${\psi}$---that can provide
some confirmation for this state ascription, and that can falsify the
state ascription (but are not guaranteed too even if ``it's wrong'').
But positive grounds for definitely believing that the system
\emph{now} (at, say, time $t$) has the property of having state
$\ket{\psi}$ cannot be had by further measurements (at times $t \ge
t'$) solely on system $B$.  This is a major way in which such quantum
``properties'' differ radically from classical ones, and, perhaps, a
reason why their change ``at distance'' via collapse, should be viewed
with less alarm than a similar classical change.  Indeed, this very
nonclassical inability to locally, retrospectively verify such
properties makes them quite different from classical properties such
as roughness and, significantly, is necessary in order to avoid
instantaneous signalling.

The way the world \emph{has} responded to Alice's probing at her site,
is correlated with the way the world \emph{will} respond at Bob's
site.  We can't use this to signal.  What's the problem?  If these
``laws of response'' are real, then this aspect of reality is not
localized, and that's about all one can say.  What's the problem?

\end{document}